\documentstyle[12pt]{article}
\advance\voffset by 1mm
\advance\hoffset by -7mm
\makeatletter
\@addtoreset{equation}{section}
\makeatother
\renewcommand{\theequation}{\thesection.\arabic{equation}}
\renewcommand{\title}[1]{\null\vspace{25mm}

\noindent{\Large{\bf #1}}\vspace{10mm}}

\newcommand{\authors}[1]{\noindent{\large #1}\vspace{3mm}

}
\newcommand{\address}[1]{\noindent #1\vspace{24mm}

}
\renewcommand{\abstract}[1]{\vspace{17mm}

\noindent{\small{\em Abstract.} #1}\vspace{2mm}

}
\newcommand {\equ}[1] {(\ref{#1})}
\newcommand{\eq}{\begin{equation}}
\newcommand{\eqn}[1]{\label{#1}\end{equation}}
\newcommand{\eea}{\end{eqnarray}}
\newcommand{\eqa}{\begin{eqnarray}}
\newcommand{\eqan}[1]{\label{#1}\end{eqnarray}}
\newcommand{\ba}{\begin{array}}
\newcommand{\ea}{\end{array}}
\newcommand{\eqac}{\begin{equation}\begin{array}{rcl}}
\newcommand{\eqacn}[1]{\end{array}\label{#1}\end{equation}}

\def\be{\begin{equation}}
\def\ee{\end{equation}}
\def\bea{\begin{eqnarray}}
\def\eea{\end{eqnarray}}
\def\bean{\begin{eqnarray*}}
\def\eean{\end{eqnarray*}}
\def\ba{\begin{array}} \def\ea{\end{array}}
\def\6{\partial} \def\a{\alpha} 
 \def\d{\delta} \def\ve{\varepsilon} 
\def\ep{\epsilon}  \def\P{\Pi} \def\F{\Phi}
 
  \def\l{\lambda}
\def\m{\mu} \def\n{\nu} \def\x{\xi} 
\def\r{\rho} \def\s{\sigma} \def\t{\tau}
 
\def\o{\omega} \def\G{\Gamma} \def\D{\Delta}
  \def\S{\Sigma}
  \def\O{\Omega}

\def\ti{\tilde} 
\def\non{\nonumber\\}
\def\={\!\!\!&=&\!\!\!}
\def\+{\!\!\!&&\!\!\!+~}
\def\-{\!\!\!&&\!\!\!-~}

\def\half{\frac{1}{2}}
\def\four{\frac{1}{4}}
\def\ix{\int d^4\!x}

\def\GO{\G^{(0)}}
\def\hb{\hbar}
\def\cb{\bar{c}}
\def\ch{\hat{c}}
\def\Fh{\hat{F}}
\def\Sh{\hat{S}}
\def\rh{\hat{\rho}}
\def\Sb{\bar{\Sigma}}
\def\lt{\left}
\def\rt{\right}
\def\h{\hat}
\renewcommand{\AA}{{\cal A}}   
     \newcommand{\DD}{{\cal D}}
     \newcommand{\FF}{{\cal F}}
\newcommand{\GG}{{\cal G}}     \newcommand{\HH}{{\cal H}}

     \newcommand{\NN}{{\cal N}}
\newcommand{\OO}{{\cal O}}     \newcommand{\PP}{{\cal P}}
     
\newcommand{\SS}{{\cal S}}     
     
\newcommand{\WW}{{\cal W}}     
     
\newcommand{\journal}[4]{{\em #1~}#2\,(19#3)\,#4;}
\newcommand{\aihp}{\journal {Ann. Inst. Henri Poincar\'e}}

\newcommand{\pr}{\journal {Phys. Rev.}}

\newcommand{\jmp}{\journal {J. Math. Phys.}}

\newcommand{\cmp}{\journal {Comm. Math. Phys.}}
\newcommand{\cqg}{\journal {Class. Quantum Grav.}}

\newcommand{\np}{\journal {Nucl. Phys.}}
\newcommand{\pl}{\journal {Phys. Lett.}}

\newcommand{\annp}{\journal {Ann. Phys. (N.Y.)}}

\begin{document}

\setcounter{page}{0}
\thispagestyle{empty}
\hspace*{\fill} REF. TUW 96-03

\title{The Yang--Mills gauge field theory in the context of a
generalized BRST--formalism including translations}

\authors{O. Moritsch\footnote{Work supported in part by the
         ``\"Osterreichische Nationalbank''
         under Contract Grant Number 5393.}, 
M. Schweda
and T. Sommer\footnote{Work supported in part by the
         ``Fonds zur F\"orderung der Wissenschaftlichen Forschung''
         under Contract Grant Number P10268-PHY.}}

\address{Institut f\"ur Theoretische Physik,
         Technische Universit\"at Wien\\
         Wiedner Hauptstra\ss e 8--10, A--1040 Wien (Austria)}
         
\begin{flushleft}
February 1996
\end{flushleft}

\abstract{We discuss the algebraic renormalization of the 
Yang--Mills gauge field theory in the presence of translations. 
Due to the translations the algebra between Sorella's
$\d$--operator, the exterior derivative and the BRST--operator
closes. Therefore, we are able to derive an 
integrated parameter formula collecting in an elegant and 
compact way all nontrivial solutions of the descent equations.}

\centerline{\Large {}}


\newpage

\section{Introduction}

In modern physics gauge field theories are essential in describing 
the properties of matter and its interactions. 
Usually, such theories are quantized using the 
BRST--formalism~\cite{brst}. 
This technique together with
the Quantum Action Principle~\cite{qap} allow for a fully
algebraic proof of their renormalizability~\cite{renorm}.
Indeed, one can calculate the invariant lagrangians
and anomalies, corresponding to a set of field transformations,
as the nontrivial solutions
of the BRST consistency condition~\cite{wz,consis}.
The latter constitutes a cohomology problem~\cite{cohom} 
due to the nilpotency of the BRST--operator.
Using the algebraic Poincar\'e lemma~\cite{poinc,descent} 
one arrives at a tower of descent equations. 

In the case of a pure Yang--Mills gauge field theory
an algebraic method for solving the descent equations has
been proposed in~\cite{decomp}. It is based on the decomposition
of the exterior spacetime derivative as a BRST--commutator
\footnote{Remark that we denote the BRST--operator in~\cite{decomp}
by $s_g$, since the letter $s$ will be reserved for the full
BRST--operator including the translations.},
\be
[\d,s_g]=d \ .
\ee
However, the algebra between the BRST--operator $s_g$, the exterior
spacetime derivative $d$ and the $\d$--operator does not close.
This leads to the presence of a further operator~\cite{decomp}, 
\be
\GG=\half[d,\d] \ ,
\label{GG_OP}
\ee 
inducing an additional tower of descent equations, which
has to be solved.

Inspired by the results of~\cite{mo1}, we show that an incorporation 
of the translations into the BRST--transformations 
leads to the disappearance of the $\GG$--operator.
Due to this fact we are able to collect the general nontrivial
solutions for the cocycles of the descent equations into
an integrated parameter formula, which will be analogously derived 
to the well--known Chern--Simons formula~\cite{chern}.

The work is organized as follows.
In section two the generalized BRST--formalism including 
translations will be introduced and
applied to pure Yang--Mills gauge field theory~\cite{tree}.
The BRST--invariance of the action defining the tree
approximation will be expressed by a Slavnov--Taylor 
identity~\cite{renorm,tree}.
Then in section three we present the functional algebra obeyed 
by the Slavnov--Taylor operator and further functional 
differential operators which appear in the constraints 
defining the tree approximation of the theory.
In section four we study the algebraic renormalization. 
We proof the stability of the action defining the tree 
approximation and the discussion of the anomaly problem 
will be done.
In section five we show that the general nontrivial solution 
for the cocycles of the descent equations corresponding to the
full BRST--operator $s$ is given in terms of
the general nontrivial solution for the
cocycles of the descent equations in the gauge sector.
Moreover, the decomposition of the exterior spacetime derivative 
as a BRST--commutator will be discussed.
Guided by Stora's derivation of the Chern--Simons formula
\cite{chern}, 
we derive an integrated parameter formula 
(see also \cite{mo1}),
which presents an elegant and compact way to 
collect all the cocycles of the descent equations in the
gauge sector in one expression.
This integrated parameter formula will be used to compute 
the gauge anomaly in four dimensions.


\section{The Yang--Mills model}

The classical dynamics of pure Yang--Mills gauge field theory 
is defined by the gauge invariant action
\footnote{As usual, Greek indices refer to the Minkowskian
space--time.}
\be
S_{inv}=-\four\,Tr\ix\,F_{\m\n}F^{\m\n} \ .
\label{CLASS_ACTION}
\ee
The field strength $F_{\m\n}$ is related to the gauge field $A_\m$ 
by the structure equation
\be
F_{\m\n}=\6_\m A_\n-\6_\n A_\m-i[A_\m,A_\n] \ .
\ee
The gauge transformation of the gauge field is given by
\be
\d_\o A_\m=D_\m\o=\6_\m\o-i[A_\m,\o] \ ,
\ee
where $D_\m$ denotes the covariant derivative. 
All fields are Lie--algebra valued, i.e. $A_{\m}=A^A_{\m}T^A$
and $F_{\m\n}=F^A_{\m\n}T^A$,
where $T^A$ are the generators of the gauge group {\bf G}
in the adjoint representation
\footnote{
Gauge group indices are denoted by capital latin letters.
The matrices $T^A$ are hermitian and traceless, and
they obey $[T^A,T^B]=if^{ABC}T^C$, $Tr(T^AT^B)=\d^{AB}$, 
where $f^{ABC}$ are the real and totally antisymmetric
structure constants of the gauge group.}.

\noindent
Moreover, the classical action \equ{CLASS_ACTION} is also invariant 
under infinitesimal translations in the Minkowskian space--time, 
\be
\d_\ve A_\m(x)=\ve^\l\6_\l A_\m(x) \ ,
\ee
with $\ve^\l$ as the infinitesimal global parameter of translations
obeying $\6_\m\ve^\n=0$.

\noindent
In order to combine the gauge invariance and the global invariance 
under translations into a single one,
we define the generalized gauge transformations according to
\be
\d:=\d_\o+\d_\ve \ .
\ee
To quantize the model we introduce the corresponding nilpotent 
BRST--operator~\cite{brst}, 
\be
s=s_g+s_T \ ,
\ee
where $s_g$ and $s_T$ are the BRST--operators corresponding
to the gauge transformations and the translations, respectively.
The latter obey the algebra 
\bea
s^2_g=s^2_T=\{s_g,s_T\}=0 \ .  
\eea
Explicitely, the several BRST-transformations are given by
\bea
sA_\m\=D_\m c + \x^\n\6_\n A^\m \ , \non 
sc\=icc + \x^\n\6_\n c \ , \non 
s\x^\m\=0 \ ,  
\label{BRST}
\eea
where $c$ is the Lie--algebra valued gauge ghost and $\x^\m$ 
is a global ghost associated to the translations, obeying
$\6_\m\x^\n=0$.

\noindent
Pure Yang--Mills gauge field theory in the tree approximation 
is described in the Landau gauge by the 
action~\cite{renorm,tree,landau}
\be
\GO=S_{inv}+S_{gf}+S_{\F\P}+S_{ext} \ ,
\ee
where
\bea
S_{inv}\=-\four\,Tr\ix\,F_{\m\n} F^{\m\n} \ , \non
S_{gf}+S_{\F\P}\=s\,Tr\ix\,\cb\, \6^\m A_\m \ , \non
S_{ext}\=s\,Tr\ix\,(-\r^\m A_\m+\s c) \ ,
\eea
are respectively the gauge invariant classical action, the sum of 
the gauge fixing term and the Faddeev--Popov term, and the source 
term. Before discussing $S_{inv}$, $S_{gf}+S_{\F\P}$, $S_{ext}$ 
and explaining the fields appearing in $\GO$, we present the 
BRST--transformations of the remaining fields:
\bea
&&s\cb=b+\x^\n\6_\n\cb~~~,~~~sb=\x^\n\6_\n b \ , \non
&&s\r^\m=\x^\n\6_\n\r^\m~~~,~~~s\s=\x^\n\6_\n\s \ , 
\label{BRST_DOUBLET}
\eea
whereby the antighost $\cb$ and the multiplier field $b$ transform
as a BRST--doublet in the gauge sector. The external sources 
$\r^\m$ and $\s$ are BRST--invariant in the gauge sector.
Due to the nilpotency of the BRST--operator the sum 
of the gauge fixing term and the Faddeev--Popov term as well as 
the source term are invariant under the BRST--transformation.

\noindent
The sum of the gauge fixing term and the Faddeev--Popov term
becomes
\be
S_{gf}+S_{\F\P}
=Tr\ix\,\lt[ b\,\6^\m A_\m-\cb\,\6^\m(D_\m c)\rt] \ .
\ee
In addition, the source term is given by
\be
S_{ext}=Tr\ix\,(\r^\m D_\m c+\s icc) \ .
\ee

\noindent
The partial derivative $\6_\m$ does not change the ghost number, 
whereas the BRST--operator raises the ghost number by one unit.
The canonical dimension and the ghost number of the gauge 
field $A_\m$, the gauge ghost $c$, the antighost $\cb$, the 
multiplier field $b$, the external sources $\r^\m$ and $\s$ and 
the global translation ghost $\x^\m$ are collected in the 
following table:
\begin{table}[h]
\begin{center}
\begin{tabular}{|c|c|c|c|c|c|c|c|}\hline
&$A_\m$&$c$&$\cb$&$b$&$\r^{\m}$&$\s$&$\x^{\m}$ \\ \hline
$dim$&1&0&2&2&3&4&-1 \\ \hline
$Q_{\F\P}$&0&1&-1&0&-1&-2&1 \\ \hline
\end{tabular} \\
\mbox{} \\
{\scriptsize  Table 1: Dimensions and Faddeev-Popov ghost 
charges of the fields.}
\end{center}
\end{table}


\section{The functional algebra}

The Slavnov--Taylor identity in the gauge sector, 
\be
\SS_g(\GO)=0 \ ,
\ee
with
\bea
\SS_g(\GO)=Tr\ix\,\lt\{\frac{\d\GO}{\d\r^\m(x)}
\frac{\d\GO}{\d A_\m(x)}
+\frac{\d\GO}{\d\s(x)}\frac{\d\GO}{\d c(x)}
+b(x)\frac{\d\GO}{\d\cb(x)}\rt\} \ ,
\eea
and the 
Ward--identity describing the invariance under translations, 
\be
\SS_T(\GO)=\x^\m\PP_\m\GO=0 \ ,
\ee 
where the generator of the translations is given by
\footnote{The sum runs over all fields $\phi$ of the model.}
\be
\PP_\m=Tr\ix\sum_{\phi}(\6_\m\phi(x))\frac{\d}{\d\phi(x)} \ ,
\ee
can be collected into a single Slavnov--Taylor identity
\be
\SS(\GO)=0 \ ,
\label{SLAVNOV_TAYLOR}
\ee
with 
\bea
\SS(\GO)\=Tr\ix\,\lt\{\lt[\frac{\d\GO}{\d\r^\m(x)}
+\x^\n\6_\n A_\m(x)\rt]\frac{\d\GO}{\d A_\m(x)}
+\lt[\x^\n\6_\n\r^\m(x)\rt]\frac{\d\GO}{\d\r^\m(x)}\rt. \non
\+\lt[\frac{\d\GO}{\d\s(x)}+\x^\n\6_\n c(x)\rt]\frac{\d\GO}{\d c(x)}
+\lt[\x^\n\6_\n\s(x)\rt]\frac{\d\GO}{\d\s(x)} \non
\+\lt.\lt[b(x)+\x^\n\6_\n\cb(x)\rt]\frac{\d\GO}{\d\cb(x)}
+\lt[\x^\n\6_\n b(x)\rt]\frac{\d\GO}{\d b(x)}\rt\} \ .
\label{SLAVNOV_TAYLOR2}
\eea
The Slavnov--Taylor identity \equ{SLAVNOV_TAYLOR}
describes the invariance of $\GO$ under the 
BRST--transformations \equ{BRST} and \equ{BRST_DOUBLET}. 

\noindent
The gauge condition is given by
\be
\frac{\d\GO}{\d b(x)}=\6^\m A_\m(x) \ ,
\label{GAUGE_CONDITION}
\ee
and the global ghost equation,
\be
\frac{\6\GO}{\6\x^\m}=0 \ ,
\ee
shows that the action defining the tree approximation does not
depend on the global translation ghost $\x^\m$.

\noindent
Moreover, one can derive a local antighost equation which controls 
the dependence of $\GO$ on the antighost $\cb$.
It can be obtained by commuting the gauge condition with the 
Slavnov--Taylor identity \equ{SLAVNOV_TAYLOR}:
\be
\GG(x)\GO=\lt\{\frac{\d}{\d\cb(x)}
+\6^\m\frac{\d}{\d\r^\m(x)}\rt\}\GO=0 \ ,
\ee
with the local antighost operator $\GG(x)$
\footnote{Remark that the local antighost operator $\GG(x)$
has nothing to do with the $\GG$--operator given in \equ{GG_OP}.}.

\noindent
In the Landau gauge~\cite{renorm,landau} one can derive an 
integrated ghost equation which controls the dependence of 
$\GO$ on the ghost field $c$. 
By calculating the functional derivative of $\GO$ 
with respect to the ghost field $c$ and using the gauge condition
one finds~\cite{renorm} 
\bea
\frac{\d\GO}{\d c(x)}+i\lt[\cb(x),\frac{\d\GO}{\d b(x)}\rt]
\=\6_\m\lt(D^\m\cb(x)+\r^\m(x)\rt) \non
\+i\lt[\r^\m(x),A_\m(x)\rt]-i\lt[\s(x),c(x)\rt] \ .
\eea
One observes that the nonlinear terms appear in a total divergence.
Therefore, an integration over spacetime will remove the 
nonlinear terms. This leads to the integrated ghost equation
\be
\HH\GO=\ix\,\lt\{\frac{\d}{\d c(x)}
+i\lt[\cb(x),\frac{\d}{\d b(x)}\rt]\rt\}\GO=\D_g \ ,
\ee
where $\HH$ denotes the integrated ghost operator and
\be
\D_g=\ix\,\lt\{i\lt[\r^\m(x),A_\m(x)\rt]-i\lt[\s(x),c(x)\rt]\rt\}
\ee
is a classical breaking, i.e. it is linear in the quantum fields.

\noindent
The invariance of the action defining the tree approximation 
under rigid gauge transformation is expressed by the Ward--identity
\be
\WW_{rig}\GO=0 \ ,
\ee
where the Ward--identity operator $\WW_{rig}$ is given by
\bea
\WW_{rig}\=\ix\,\lt(i\lt[A_\m(x),\frac{\d}{\d A_\m(x)}\rt]+
i\lt\{\r^\m(x),\frac{\d}{\d\r^\m(x)}\rt\}\rt. \non
\+i\lt\{c(x),\frac{\d}{\d c(x)}\rt\}+
i\lt[\s(x),\frac{\d}{\d\s(x)}\rt] \non
\+\lt.i\lt\{\cb(x),\frac{\d}{\d\cb(x)}\rt\}+
i\lt[b(x),\frac{\d}{\d b(x)}\rt]\rt) \ .
\eea

\noindent
The Slavnov--Taylor operator acting on an 
arbitrary functional $\FF$ with even ghost charge is 
\bea
\SS(\FF)\=Tr\ix\,\lt\{
\lt[\frac{\d\FF}{\d\r^\m(x)}+\x^\n\6_\n A_\m(x)\rt]
\frac{\d\FF}{\d A_\m(x)}+
\lt[\x^\n\6_\n\r^\m(x)\rt]\frac{\d\FF}{\d\r^\m(x)}\rt. \non
\+\lt.\lt[\frac{\d\FF}{\d\s(x)}+\x^\n\6_\n c(x)\rt]
\frac{\d\FF}{\d c(x)}+
\lt[\x^\n\6_\n\s(x)\rt]\frac{\d\FF}{\d\s(x)}\rt.  \non
\+\lt.\lt[b(x)+\x^\n\6_\n\cb(x)\rt]\frac{\d\FF}{\d\cb(x)}+
\lt[\x^\n\6_\n b(x)\rt]\frac{\d\FF}{\d b(x)}\rt\} \ ,
\eea
and the linearized Slavnov--Taylor operator 
can be written as
\bea
\SS_{\FF}\=Tr\ix\,\lt\{
\lt[\frac{\d\FF}{\d\r^\m(x)}+\x^\n\6_\n A_\m(x)\rt]
\frac{\d}{\d A_\m(x)}+\lt[\frac{\d\FF}{\d A_\m(x)}
+\x^\n\6_\n\r^\m(x)\rt]\frac{\d}{\d\r^\m(x)}\rt. \non
\+\lt.\lt[\frac{\d\FF}{\d\s(x)}+\x^\n\6_\n c(x)\rt]
\frac{\d}{\d c(x)}+\lt[\frac{\d\FF}{\d c(x)}+\x^\n\6_\n\s(x)\rt]
\frac{\d}{\d\s(x)}\rt. \non
\+\lt.\lt[b(x)+\x^\n\6_\n\cb(x)\rt]\frac{\d}{\d\cb(x)}+
\lt[\x^\n\6_\n b(x)\rt]\frac{\d}{\d b(x)}\rt\} \ .
\eea

\noindent
The functional algebra which is valid for any functional $\FF$ 
with even ghost charge is given by the following relations:
\begin{itemize}
\item{The nilpotency of the Slavnov--Taylor operator 
is contained in the two following identities:
\bea
\SS_{\FF}\lt(\SS(\FF)\rt)\=0 \ , \label{LIN_SLAV} \\
\SS(\FF)\=0~~~~~\Longrightarrow~~~~~\SS_{\FF}\SS_{\FF}=0 \ .
\label{LIN_SLAV_2}
\eea
}
\item{Commuting the partial derivative with respect to the global 
translation ghost with the Slavnov--Taylor operator, 
one gets
\bea
\frac{\6}{\6\x^\m}\,\SS(\FF)+
\SS_{\FF}\lt(\frac{\6\FF}{\6\x^\m}\rt)\=\PP_\m\FF \ , \non[3mm]
\PP_\m\SS(\FF)-\SS_{\FF}(\PP_\m\FF)\=0 \ .  
\label{PS_COMM}
\eea
}
\item{Commuting the gauge condition with the  
Slavnov--Taylor operator, one obtains
\bea
\frac{\d}{\d b(x)}\,\SS(\FF)-
\SS_{\FF}\lt(\frac{\d\FF}{\d b(x)}-\6^\m A_\m(x)\rt)
\=\GG(x)\FF-\x^\n\6_\n\lt(\frac{\d\FF}{\d b(x)}
-\6^\m A_\m(x)\rt), \non[3mm]
\GG(x)\SS(\FF)+\SS_{\FF}\lt[\GG(x)\FF\rt]
\=\x^\n\6_\n\lt[\GG(x)\FF\rt] \ .
\eea
}
\item{Commuting the integrated ghost operator with the 
Slavnov--Taylor operator, gives
\bea
\HH\SS(\FF)+\SS_{\FF}(\HH\FF-\D_g)\=\WW_{rig}\FF \ , \non[3mm]
\WW_{rig}\SS(\FF)-\SS_{\FF}(\WW_{rig}\FF)\=0 \ .
\label{WS_COMM}
\eea
}
\end{itemize}


\section{Renormalization, stability and anomalies}

The aim of the renormalization is to construct an extension of the
theory at the tree level to all orders of perturbation theory.
This extension will be described by the vertex functional,
\be
\G=\GO+\OO(\hb) \ ,
\ee
generating the 1PI Green functions.
One has to examine whether this vertex functional obeys a 
Slavnov--Taylor identity, being nonlinear in $\G$,
\be
\SS(\G)=0 \ .
\label{SLAV_TAYL}
\ee
Since the constraints being linear in the quantum
fields are renormalizable to all orders of perturbation theory, the
following relations are valid~\cite{renorm,landau}:
\begin{itemize}
\item{the gauge condition,
\be
\frac{\d\G}{\d b(x)}=\6^\m A_\m(x) \ ,
\label{GAUGE_CON}
\ee
}
\item{the integrated ghost equation,
\be
\HH\G=\D_g \ ,
\ee
}
\item{the global ghost equation,
\be
\frac{\6\G}{\6\x^\m}=0 \ ,
\ee
}
\item{the local antighost equation,
\be
\GG(x)\G=0 \ ,
\ee
}
\item{the invariance under rigid gauge transformations,
\be
\WW_{rig}\G=0 \ ,
\ee
}
\item{the invariance under translations,
\be
\SS_T(\G)=\x^\m\PP_\m\G=0 \ .
\label{TRANS}
\ee
}
\end{itemize}
Within the framework of the algebraic renormalization procedure
\cite{renorm},
based on the general grounds of power counting and locality, the
discussion of the extension of the theory in the tree approximation
to all orders of perturbation theory
is organized according to two independent parts:
First, the study of the stability of the classical action under
radiative corrections.
This amounts to find the invariant counterterms and to check if they
all correspond to a renormalization of the free parameters of the
classical theory.
Second, the search for anomalies, i.e. the investigation whether 
the symmetries of the theory survive in the presence of radiative
corrections.

\subsection{Stability}

In order to check that the action in the tree approximation is 
stable under radiative corrections, one perturbs it by an 
arbitrary integrated local functional $\S_c$,
\bea
\S=\GO+\a\S_c~~~,~~~\a\S_c=\OO(\hb) \ ,
\label{STAB}
\eea
where $\a$ is an infinitesimal parameter with vanishing canonical
dimension and vanishing ghost number.
The functional $\S_c$ has the same quantum numbers as the action 
in the tree approximation\footnote{One has $dim(d^4x)=-4$.}.

\noindent
One requires that the perturbed action $\S$ satisfies the same
constraints defining the theory at the tree level, i.e.
\equ{SLAV_TAYL}--\equ{TRANS}.
The perturbation $\S_c$ a priori depends on all fields and on the
parameter $\x^\m$,
\be
\S_c=\S_c[A_\m,\,c,\,\cb,\,b,\,\r^\m,\,\s](\x^\m) \ .
\ee
The gauge condition, the integrated ghost equation 
and the global ghost equation together with the same 
constraints for the action at the tree level imply
\bea
\frac{\d\S_c}{\d b(x)}=0~~~,~~~\HH\S_c=0~~~,~~~
\frac{\6\S_c}{\6\x^\m}=0 \ .
\label{b_INV}
\eea
Therefore, the perturbation $\S_c$ does not depend on the 
multiplier field $b(x)$ and the global translation ghost 
$\x^\m$, i.e.
\be
\S_c=\S_c[A_\m,\,c,\,\cb,\,\r^\m,\,\s] \ .
\ee
Since
\be
\ti{\HH}\S_c=\ix\,\frac{\d\S_c}{\d c(x)}=0 \ ,
\ee
the dependence on the ghost $c(x)$ has to be a total divergence.
The local antighost equation, the invariance under rigid gauge 
transformations and the invariance under translations together 
with the same constraints for the action at the tree level imply
\bea
\GG(x)\S_c=0~~~,~~~\WW_{rig}\S_c=0~~~,~~~
\PP_\m\S_c=0 \ . 
\label{W_SIGMA}
\eea
From eqs.\equ{b_INV} and \equ{W_SIGMA} follows
\bea
\ti{\WW}_{rig}\S_c\=\ix
\lt(i\lt[A_\m(x),\frac{\d\S_c}{\d A_\m(x)}\rt]+
i\lt\{\r^\m(x),\frac{\d\S_c}{\d\r^\m(x)}\rt\}+
i\lt\{c(x),\frac{\d\S_c}{\d c(x)}\rt\}\rt. \non
\+\lt.i\lt[\s(x),\frac{\d\S_c}{\d\s(x)}\rt]+
i\lt\{\cb(x),\frac{\d\S_c}{\d\cb(x)}\rt\}\rt)=0 \ , \\[3mm]
\ti{\PP}_\m\S_c\=Tr\ix
\lt[\6_\m A_\n(x)\,\frac{\d\S_c}{\d A_\n(x)}+
\6_\m\r^\n(x)\,\frac{\d\S_c}{\d\r^\n(x)}+
\6_\m c(x)\,\frac{\d\S_c}{\d c(x)}\rt. \non
\+\lt.\6_\m\s(x)\,\frac{\d\S_c}{\d\s(x)}+
\6_\m\cb(x)\,\frac{\d\S_c}{\d\cb(x)}\rt]=0 \ .
\eea
The local antighost equation $\GG(x)\S_c=0$ suggests the following 
change of field variables
\bea
\rh^\m(x)=\r^\m(x)+\6^\m\cb(x)~~~,~~~\h{\cb}(x)=\cb(x) \ ,
\eea
with
\be
\S_c[A_\m,\,c,\,\cb,\,\r^\m,\,\s]=
\Sb_c[A_\m,\,c,\,\h{\cb},\,\rh^\m,\,\s] \ .
\ee
Then the local antighost equation becomes
\be
\h{\GG}(x)\Sb_c=\frac{\d\Sb_c}{\d\h{\cb}(x)}=0 \ .
\ee
Therefore, the perturbation $\Sb_c$ depends on $\r^\m$ and
$\cb$ only through the combination $\rh^\m=\r^\m+\6^\m\cb$, i.e.
\be
\Sb_c=\Sb_c[A_\m,\,c,\,\rh^\m,\,\s] \ .
\ee

\noindent
Applying the Slavnov--Taylor operator to the 
perturbed action, one gets
\be
\SS(\S)=\SS(\GO+\a\S_c)=\SS(\GO)+\a\,\SS_{\GO}(\S_c)+\OO(\a^2) \ .
\label{PERT}
\ee
Using the Slavnov--Taylor identity for the action in the tree 
approximation \equ{SLAVNOV_TAYLOR}, the Slavnov--Taylor identity
imposed to the perturbed action \equ{STAB} translates at the 
first order in $\a$ into the following condition on the 
perturbation $\S_c$:
\be
\SS_{\GO}(\S_c)=0 \ .
\ee
This equation is the BRST consistency condition in the 
ghost number sector zero.
It constitutes a cohomology problem, due to the nilpotency of the 
linearized Slavnov--Taylor operator,
\be
\SS_{\GO}\SS_{\GO}=0 \ ,
\ee
which follows from the validity of the Slavnov--Taylor identity 
for the action in the tree approximation \equ{SLAVNOV_TAYLOR},
where \equ{LIN_SLAV_2} has been used.
The solution of the BRST consistency condition in the 
ghost number sector zero can always be written as the sum of a 
trivial cocycle $\SS_{\GO}\h{\S}$, where $\h{\S}$ has ghost 
number $-1$, and the nontrivial elements belonging to the 
cohomology of $\SS_{\GO}$ in the ghost number sector zero, i.e. 
which cannot be written as $\SS_{\GO}$--variations:
\be
\S_c=\S_{ph}+\SS_{\GO}\h{\S} \ .
\ee
The trivial cocycle $\SS_{\GO}\h{\S}$ corresponds to field
renormalizations which are unphysical.
Using the functional derivatives of $\GO$ with respect to the 
gauge field $A_\m$, the 
gauge ghost $c$ and the classical external sources $\r^\m$ 
and $\s$, the invariance of $\Sb_c$ under translations 
and since $\d\Sb_c/\d b(x)=0$, 
the BRST consistency condition reads in the new variables
\bea
\SS_{\GO}(\Sb_c)\=Tr\ix\lt[
D_\m c(x)\,\frac{\d\Sb_c}{\d A_\m(x)}+
\lt(D_\n F^{\n\m}(x)+i\lt\{c(x),\rh^\m(x)\rt\}\rt)
\frac{\d\Sb_c}{\d\rh^\m(x)}\rt.  \non
\+\lt.ic(x)c(x)\frac{\d\Sb_c}{\d c(x)}+
\lt(D_\m\rh^\m(x)+i\lt[c(x),\s(x)\rt]\rt)
\frac{\d\Sb_c}{\d\s(x)}\rt]=0 \ .
\eea
Therefore, the perturbation $\Sb_c=\Sb_c[A_\m,\,c,\,\rh^\m,\,\s]$
is an integrated local functional with canonical dimension 
zero and ghost number zero obeying the following set 
of constraints:
\begin{itemize}
\item{the integrated ghost equation,
\be
\h{\HH}\Sb_c=\ix\,\frac{\d\Sb_c}{\d c(x)}=0 \ ,
\label{INT_GHOST_EQU}
\ee
}
\item{the invariance under rigid gauge transformations,
\bea
\h{\WW}_{rig}\Sb_c\=\ix\lt(
i\lt[A_\m(x),\frac{\d\Sb_c}{\d A_\m(x)}\rt]+
i\lt\{\rh^\m(x),\frac{\d\Sb_c}{\d\rh^\m(x)}\rt\}\rt.  \non
\+\lt.i\lt\{c(x),\frac{\d\Sb_c}{\d c(x)}\rt\}+
i\lt[\s(x),\frac{\d\Sb_c}{\d\s(x)}\rt]\rt)=0 \ ,
\eea
}
\item{the invariance under translations,
\bea
\h{\PP}_\m\Sb_c\=Tr\ix\lt[
\6_\m A_\n(x)\,\frac{\d\Sb_c}{\d A_\n(x)}+
\6_\m\rh^\n(x)\,\frac{\d\Sb_c}{\d\rh(x)}\rt.   \non
\+\lt.\6_\m c(x)\,\frac{\d\Sb_c}{\d c(x)}+
\6_\m\s(x)\,\frac{\d\Sb_c}{\d\s(x)}\rt]=0 \ ,
\eea
}
\item{and the BRST--consistency condition,
\bea
\SS_{\GO}(\Sb_c)\=Tr\ix\lt[c(x)\DD(x)\Sb_c
+D_\n F^{\n\m}(x)\,\frac{\d\Sb_c}{\d\rh^\m(x)}\rt. \non
\+\lt.D_\m\rh^\m(x)\,\frac{\d\Sb_c}{\d\s(x)}\rt]=0 \ ,
\label{BRST_CC}
\eea
where the abbreviation
\bea
\DD(x)\=-D_\m\frac{\d}{\d A_\m(x)}
+i\lt\{\rh^\m(x),\frac{\d}{\d\rh^\m(x)}\rt\} \non
\+\frac{i}{2}\lt\{c(x),\frac{\d}{\d c(x)}\rt\}
+i\lt[\s(x),\frac{\d}{\d\s(x)}\rt]
\label{ABBRE}
\eea
has been used.
}
\end{itemize}
It will be shown in the appendix that the solution of the set of 
constraints \equ{INT_GHOST_EQU}--\equ{BRST_CC} is given by
\be
\Sb_c=-k\,\four\,Tr\ix\,F_{\m\n}(x)F^{\m\n}(x) \ ,
\label{COUNTER_TERM}
\ee
which is the most general nontrivial perturbation of the action
in the tree approximation.
The perturbation $\Sb_c$ depends on the parameter $k$ which 
corresponds to a possible multiplicative renormalization of the 
gauge coupling constant $g$ being set in the whole work equal 
to one. This is an algebraic result which just shows that the 
nontrivial invariant counterterm \equ{COUNTER_TERM} 
can be reabsorbed into the action at the tree
level by a renormalization of its coefficient.
It means that no new terms appear at the $n$--loop levels with
$n\geq 1$. Therefore, the action in the tree approximation 
is stable.

\subsection{Anomalies}

In the following we investigate if the BRST--symmetry is
preserved in the presence of radiative corrections.
Indeed, the aim of the renormalization procedure is to examine if it
is possible to define a vertex functional, $\G=\GO+\OO(\hb)$,
obeying as in the tree approximation the set of 
constraints \equ{SLAV_TAYL}--\equ{TRANS}.
Since the operators in \equ{GAUGE_CON}--\equ{TRANS}
are linear differential operators and 
since the breakings are linear in the quantum fields one can 
assume the validity of the equations \equ{GAUGE_CON}--\equ{TRANS}
at the full quantum level, i.e. to all orders of 
perturbation theory.

\noindent
Actually the program will fail because the nonlinear 
Slavnov--Taylor identity \equ{SLAV_TAYL} 
will turn out to be anomalous
\be
\SS(\G)=r\D_{AB} \ ,
\ee
where $\D_{AB}$ is the anomaly to be derived in the following and $r$ 
is a well--known function of order $\hb$ of the coupling constant $g$, 
which however cannot be determined by the pure algebraic method used 
here. The search for the breaking $\D_{AB}$ of the  
Slavnov--Taylor identity requires some care. 

\noindent
The breaking is controlled by the quantum action 
principle, which implies
\be
\SS(\G)=0+\hb\,\D\cdot\G \ ,
\ee
where the quantum breaking $\D\cdot\G$ is, at lowest order in $\hb$,
an integrated local functional
\bea
\D\cdot\G=\D+\OO(\hb\D)~~~,~~~\D=\ix\,\D(x) \ ,
\eea
with canonical dimension zero and Faddeev--Popov charge one.
Thus one gets
\be
\SS(\G)=0+\hb\D+\OO(\hb^2) \ .
\label{BROKEN_SLAV}
\ee
Applying the linear functional differential operator $\SS_\G$
and using the algebraic relation \equ{LIN_SLAV} 
one gets
\be
0=0+\hb\,\SS_\G\D+\OO(\hb^2) \ .
\label{CC_M}
\ee
From $\G=\GO+\OO(\hb)$ follows
\be
\SS_\G=\SS_{\GO}+\OO(\hb) \ ,
\ee
and eq.\equ{CC_M} becomes
\be
0=0+\hb\,\SS_{\GO}\D+\OO(\hb^2) \ .
\ee
Therefore, the integrated local functional $\D$ obeys the  
BRST consistency condition in the ghost number sector one,
\be
\SS_{\GO}\D=0 \ .
\label{ANOM_CC}
\ee

\noindent
The functional algebra \equ{PS_COMM}--\equ{WS_COMM}, 
written for the functional $\G$,
the renormalized constraints \equ{GAUGE_CON}--\equ{TRANS},
and the broken Slavnov--Taylor identity \equ{BROKEN_SLAV}
lead to the following set of constraints which the breaking,
\be
\D=\D[A_\m,\,c,\,\cb,\,b,\,\r^\m,\,\s](\x^\m) \ ,
\ee
has to obey:
\bea
\frac{\d\D}{\d b(x)}\=0 \ , \label{GC} \\
\HH\D\=0 \ , \label{H_DELTA} \\
\frac{\6\D}{\6\x^\m}\=0 \ , \label{GG} \\
\GG(x)\D\=0 \ , \label{LAE} \\
\WW_{rig}\D\=0 \ , \\
\PP_\m\D\=0 \ .
\eea
The constraints \equ{GC} and \equ{GG} imply that the breaking 
$\D$ does not depend on the multiplier field $b(x)$ and the 
global translation ghost $\x^\m$,
\be
\D=\D[A_\m,\,c,\,\cb,\,\r^\m,\,\s] \ .
\ee
The integrated ghost equation \equ{H_DELTA} becomes
\be
\h{\HH}\D=\ix\,\frac{\d\D}{\d c(x)}=0 \ ,
\ee
stating that the dependence of the breaking $\D$ on the ghost $c(x)$
has to be a total divergence.
The local antighost equation \equ{LAE} will be transformed by the 
following change of field variables,
\bea
\rh^\m(x)=\r^\m(x)+\6^\m\cb(x)~~~,~~~\h{\cb}(x)=\cb(x) \ ,
\label{NEW}
\eea
into the equation
\be
\h{\GG}(x)\D=\frac{\d\D}{\d\h{\cb}(x)}=0 \ .
\ee
This implies that the breaking $\D$ depends on $\r^\m$ and $\cb$
only through the combination $\rh^\m=\r^\m+\6^\m\cb$, i.e.
\be
\D=\D[A_\m,\,c,\,\rh^\m,\,\s] \ .
\ee
Furthermore, the breaking $\D$ has to be invariant under the 
rigid gauge transformations and translations.
The BRST consistency condition,
\be
\SS_{\GO}\D=0 \ ,
\label{GEN_CC}
\ee
constitutes a cohomology problem in the space of integrated local 
functionals with dimension zero and ghost number one due to the
nilpotency of the linearized Slavnov--Taylor operator 
$\SS_{\GO}$, i.e.
\be
\SS_{\GO}\SS_{\GO}=0 \ ,
\ee
which follows from the validity of the Slavnov--Taylor 
identity \equ{SLAVNOV_TAYLOR}.
The solution of the BRST consistency condition \equ{GEN_CC} 
can always be written as the sum of a trivial cocycle 
$\SS_{\GO}\h{\D}$, where $\h{\D}$ has ghost number zero, and the 
nontrivial elements belonging to the cohomology of $\SS_{\GO}$ in 
the ghost number sector one, i.e. which cannot be written as 
$\SS_{\GO}$--variations,
\be
\D=\SS_{\GO}\h{\D}+\D^\ast \ .
\ee
The trivial cocycle $\SS_{\GO}\h{\D}$ can be absorbed into the 
vertex functional $\G$ as an integrated local noninvariant 
counterterm $-\hb\h{\D}$,
\be
\G~~~\longrightarrow~~~\G-\hb\h{\D} \ ,
\label{ABSORB}
\ee
leading to
\bea
\SS(\G-\hb\h{\D})=\SS(\G)-\hb\,\SS_\G\h{\D}+\OO(\hb^2)
=\hb\D^\ast+\OO(\hb^2) \ .
\eea
The construction of the explicit form of the anomaly $\D^\ast$ is
explained in the final part of this subsection.
The BRST consistency condition in the ghost number 
sector one becomes in the new variables \equ{NEW} 
\bea
\SS_{\GO}\D\=Tr\ix\lt[c(x)\DD(x)\D
+D_\n F^{\n\m}(x)\,\frac{\d\D}{\d\rh^\m(x)}\rt. \non
\+\lt.D_\m\rh^\m(x)\,\frac{\d\D}{\d\s(x)}\rt]=0 \ ,
\label{CC_3}
\eea
with the abbreviation \equ{ABBRE}.
Before solving the consitency condition \equ{CC_3}, 
i.e. deriving the anomaly $\D^\ast$, one eliminates
the dependence of $\D$ on the external sources $\rh^\m$ and $\s$.
The most general dependence on $\s$ which is compatible with 
dimension and ghost number is~\cite{renorm}
\be
\D=l_1\,Tr\ix\,\s(x)c(x)c(x)c(x)+\ldots \ ,
\ee
where $l_1$ is an arbitrary constant and the low dots denote the 
terms which are independent of $\s$.
Using
\bea
\SS_{\GO}\s(x)\=i\lt[c(x),\s(x)\rt]+\ldots \ , \non
\SS_{\GO}c(x)\=ic(x)c(x) \ ,
\eea
the consistency condition yields
\be
0=\SS_{\GO}\D=-i\,l_1\,Tr\ix\,\s(x)c^4(x)+\ldots \ ,
\ee
implying
\be
l_1=0 \ .
\ee
Therefore, $\D$ is independent of $\s$,
\be
\frac{\d\D}{\d\s(x)}=0 \ .
\label{SIGMA_INV}
\ee
The most general dependence on $\rh^\m$ which is compatible with
dimension and ghost number is given by~\cite{renorm}
\be
\D=Tr\ix\,\rh^\m(x)R_\m(A,c)(x)+\ldots \ ,
\label{DELTA_3}
\ee
where now the low dots denote the terms which are independent 
of $\rh^\m$ and $R_\m(A,c)$ is the most general polynomial 
depending on $A_\m$ and $c$ with canonical dimension one and 
ghost number two
\bea
R_\m(A,c)(x)\=l_2\6_\m c(x)c(x)+l_3c(x)\6_\m c(x)
+l_4A_\m(x)c(x)c(x) \non
\+l_5c(x)A_\m(x)c(x)+l_6c(x)c(x)A_\m(x) \ .
\eea
The constants $l_2$, $l_3$, $l_4$, $l_5$ and $l_6$ are arbitrary.
In order to restrict the coefficients 
$l_2$, $l_3$, $l_4$, $l_5$, $l_6$
one uses the BRST consistency condition \equ{CC_3}
and the fact that the $\s$--dependence has been already eliminated,
\equ{SIGMA_INV}.
Writing only the terms depending on $\rh^\m$, the l.h.s of the 
consistency condition becomes
\bea
\SS_{\GO}\D\=Tr\ix\,\lt\{(-l_4-il_2)\rh^\m(x)\6_\m c(x)c(x)c(x)+
l_5\rh^\m(x)c(x)\6_\m c(x)c(x)\rt. \non
\+\lt.(-l_6+il_3)\rh^\m(x)c(x)c(x)\6_\m c(x)-
il_5\rh^\m(x)c(x)A_\m(x)c(x)c(x)\rt.  \non
\+\lt.il_5\rh^\m(x)c(x)c(x)A_\m(x)c(x)\rt\}+\ldots \ .
\eea
The consistency condition \equ{CC_3} implies the following 
relations between the constants $l_2$, $l_3$, $l_4$, $l_5$, $l_6$:
\bea
l_4=-il_2~~~,~~~l_5=0~~~,~~~l_6=il_3 \ ,
\eea
and $R_\m(A,c)$ becomes
\bea
R_\m(A,c)(x)\=l_2\6_\m c(x)c(x)+l_3c(x)\6_\m c(x)
-il_2A_\m(x)c(x)c(x) \non
\+il_3c(x)c(x)A_\m(x) \ .
\eea
Therefore, the most general dependence on $\rh^\m$ is given by
\bea
\D\=Tr\ix\,\rh^\m(x)\lt[l_2\6_\m c(x)c(x)+l_3c(x)\6_\m c(x)\rt. \non
\-\lt.il_2A_\m(x)c(x)c(x)+il_3c(x)c(x)A_\m(x)\rt]+\ldots \ .
\eea
Moreover, it follows that $\D$ can be written as the 
$\SS_{\GO}$--variation of
\be
\h{\D}=Tr\ix\,\lt[-l_2\rh^\m(x)A_\m(x)c(x)
+l_3\rh^\m(x)c(x)A_\m(x)\rt] \ ,
\ee
up to terms independent of $\rh^\m$ which are denoted by the 
low dots,
\be
\SS_{\GO}\h{\D}=\D+\ldots \ .
\ee
Hence the $\rh^\m$--dependence of the breaking $\D$ is trivial and it
can be absorbed into $\G$ according to eq.\equ{ABSORB}, 
which leads to
\be
\frac{\d\D}{\d\rh^\m(x)}=0 \ .
\label{R_HAT_INV}
\ee
From eqs.\equ{SIGMA_INV} and \equ{R_HAT_INV} 
follow that the breaking $\D$ is independent of $\rh^\m$ and $\s$.
Therefore, it does only depend on $A_\m$ and $c$,
\be
\D=\D[A_\m,c] \ .
\label{BREAK}
\ee

\noindent
Finally, let us discuss the well--known derivation of the 
breaking \equ{BREAK}. It will be shown that it 
is equal to the gauge anomaly in four dimensions.
The breaking \equ{BREAK} is the general nontrivial solution 
of the BRST consistency condition,
\be
\SS_{\GO}\D=Tr\ix\,\lt[D_\m c(x)\,\frac{\d\D}{\d A_\m(x)}
+ic(x)c(x)\,\frac{\d\D}{\d c(x)}\rt]=0 \ .
\label{C1}
\ee
Moreover, it obeys the integrated ghost equation,
\be
\ix\,\frac{\d\D}{\d c(x)}=0 \ ,
\ee
the Ward--identity describing the invariance under rigid gauge
transformations,
\be
\ix\,\lt(i\lt[A_\m(x),\frac{\d\D}{\d A_\m(x)}\rt]
+i\lt\{c(x),\frac{\d\D}{\d c(x)}\rt\}\rt)=0 \ ,
\ee
and the Ward--identity describing the invariance under translations,
\be
Tr\ix\,\lt[\6_\m A_\n(x)\frac{\d\D}{\d A_\n(x)}
+\6_\m c(x)\frac{\d\D}{\d c(x)}\rt]=0 \ .
\label{C4}
\ee

\noindent
The general nontrivial solution of the set of constraints
\equ{C1}--\equ{C4} is given by the Adler--Bardeen 
nonabelian gauge anomaly, $\D=r\AA$,
\be
\AA=Tr\ix\,\ep^{\m\n\s\t}\,c(x)\,\6_\m
\lt[\6_\n A_\s(x)A_\t(x)+\frac{i}{2}A_\n(x)A_\s(x)A_\t(x)\rt] \ .
\label{ADLER}
\ee
The actual presence of the gauge anomaly depends on the
nonvanishing of its coefficient $r$, which cannot be determined by  
the algebraic renormalization~\cite{renorm}.

\noindent
The explicit calculation of \equ{ADLER} is presented in~\cite{decomp},
where an operator $\d$, which allows to decompose the
exterior derivative,
\be
[\d,s_g]=d \ ,
\ee
is introduced. In~\cite{decomp} the algebra between the 
operators $s_g$, $d$ and $\d$ does not close.
This leads to the presence of a further operator,
\be
\GG=\half[d,\d] \ ,
\ee
inducing an additional tower of descent equations which has to be
solved in order to compute the general nontrivial solution of the
descent equations~\cite{decomp}.

\noindent
In the next section an alternative algebraic method to solve the 
BRST consistency condition will be presented.
It is based on the use of the full BRST--operator,
\be
s=s_g+s_T \ ,
\ee
including the translations, which allow for avoiding the 
operator $\GG$. This will be possible due to the fact that the
$s$--cohomology is isomorphic to that of $s_g$ up to
trivial contributions.


\section{The BRST consistency condition}

The renormalization procedure discussed before has led to find the 
general nontrivial solution of the BRST consistency condition,
\be
s_g\D=0 \ ,
\label{BRST_CONSISTENCY_CONDITION}
\ee
which constitutes a cohomology problem due to the nilpotency of the 
BRST operator in the gauge sector,
\be
s^2_g=0 \ .
\ee
The integrated local functional $\D$ turned out to depend on the 
gauge field $A_\m$ and the ghost field $c$ only, $\D=\D[A_\m,c]$.
As proven by~\cite{wz,consis},
the use of the calculus of differential forms is no 
restriction to the generality of the solution of the BRST 
consistency condition \equ{BRST_CONSISTENCY_CONDITION}.
Writing
\be
\D=\int Q^G_4 \ ,
\ee
where\footnote{
The upper index denotes the ghost number, whereas the lower 
index denotes the form degree.}
 $Q^G_4=\D^G(x)d^4x$
is a volume form, the BRST consistency condition 
\equ{BRST_CONSISTENCY_CONDITION}
translates into the local equation
\be
s_gQ^G_4+dQ^{G+1}_3=0 \ ,
\label{LOCAL_EQU}
\ee
with $d=dx^\m\6_\m$ the nilpotent exterior derivative.
In the ghost number sector zero $(G=0)$, a nontrivial solution 
for $Q^0_4$, $Q^0_4\neq d\h{Q}^0_3$,
represents an invariant Lagrangian.
Since $c$ has ghost number one, $\D$ is only a functional 
of $A_\m$.
The solution for $\D$ has been already given in the previous 
section \equ{COUNTER_TERM}.

\noindent
In the ghost number sector one $(G=1)$, a nontrivial solution 
for $Q^1_4$, $Q^1_4\neq s\h{Q}^0_4+d\h{Q}^1_3$,
represents a possible canditate for an anomaly.
The local equation \equ{LOCAL_EQU} 
becomes
\be
s_gQ^1_4+dQ^2_3=0 \ .
\ee
It is a cohomology problem with respect to $s_g$ modulo $d$.
Using the algebra, $s^2_g=0$, $d^2=0$, 
$\{s_g,d\}=0$, and the algebraic Poincar\'e lemma~\cite{poinc}
one gets the tower of descent equations,
\bea
&&s_gQ^1_4+dQ^2_3=0 \ , \non
&&s_gQ^2_3+dQ^3_2=0 \ , \non
&&s_gQ^3_2+dQ^4_1=0 \ , \non
&&s_gQ^4_1+dQ^5_0=0 \ , \non
&&s_gQ^5_0=0 \ .
\label{TOWER}
\eea
The last equation of the tower is a local cohomology 
problem with respect to $s_g$.

\noindent
In order to derive an alternative algebraic method for solving the 
descent equations \equ{TOWER},
the relation between the solution of the $s$ modulo $d$ cohomology
and the solution of the $s_g$ modulo $d$ cohomology 
has to be analyzed.


\subsection{The cohomology}

In order to be quite general,
the BRST consistency condition corresponding to the full 
BRST--operator,
\be
s\Xi=0 \ ,
\ee  
will be discussed in a spacetime with $N$ dimensions.
The integrated local functional $\Xi=\int\O^0_N$ is assumed to 
depend on $A_\m$, $c$ and $\x^\m$, 
\be
\Xi=\Xi[A_\m,c](\x^\m) \ ,
\ee
where $\x^\m$ plays the role of a parameter.
The corresponding descent equations are
\bea
&&s\O^0_N+d\O^1_{N-1}=0 \ , \non
&&~~~~~~\cdots  \non
&&s\O^{N-1}_1+d\O^N_0=0 \ , \non
&&s\O^N_0=0 \ .
\label{TOWER-2}
\eea
The cocycles $\O^k_{N-k}$ can be expanded as series in 
powers of $\x^\m$.
The expansion of the BRST operator as a series in
powers of $\x^\m$ is
\be
s=s_g+s_T=s_g+\x^\m\6_\m \ .
\ee
Introducing the counting operator
\be
\NN_\x = \x^\m\frac{\6}{\6\x^\m} \ ,
\ee
one gets
\be
\O^k_{N-k}=\sum^k_{m=0}(\O^k_{N-k})_{(m)}~~~,~~~k=0,\ldots,N \ ,
\ee
with
\be
\NN_\x(\O^k_{N-k})_{(m)}=m(\O^k_{N-k})_{(m)}~~~,~~~0\leq m\leq k~~~,
~~~k=0,\ldots,N \ .
\ee
Moreover, one has
\be
s=s_{(0)}+s_{(1)}=s_g+s_T \ ,
\ee
with $[\NN_\x,s_{(0)}]=0$ and $[\NN_\x,s_{(1)}]=s_{(1)}$.
The descent equations \equ{TOWER-2} become
\bea
&&(s_{(0)}+s_{(1)})\sum^k_{m=0}(\O^k_{N-k})_{(m)}+
d\sum^{k+1}_{m=0}(\O^{k+1}_{N-k-1})_{(m)}=0~~~,~~~
k=0,\ldots,N-1 \ , \non
&&(s_{(0)}+s_{(1)})\sum^N_{m=0}(\O^N_0)_{(m)}=0 \ .
\label{TOW}
\eea
Arranging then the terms with respect to increasing power in 
$\x^\m$, one obtains
\bea
&&s_{(0)}(\O^0_N)_{(0)}+d(\O^1_{N-1})_{(0)} 
+s_{(1)}(\O^0_N)_{(0)}+d(\O^1_{N-1})_{(1)}=0 \ , \non[3mm]
&&s_{(0)}(\O^k_{N-k})_{(0)}+d(\O^{k+1}_{N-k-1})_{(0)} \non
&&~~~+~\sum^k_{l=1}\lt[s_{(0)}(\O^k_{N-k})_{(l)}
+s_{(1)}(\O^k_{N-k})_{(l-1)}+d(\O^{k+1}_{N-k-1})_{(l)}\rt] \non
&&~~~+~s_{(1)}(\O^k_{N-k})_{(k)}+d(\O^{k+1}_{N-k-1})_{(k+1)}=0~~~,
~~~k=1,\ldots,N-1 \ , \non[3mm]
&&s_{(0)}(\O^N_0)_{(0)} 
+\sum^N_{l=1}[s_{(0)}(\O^N_0)_{(l)}+s_{(1)}(\O^N_0)_{(l-1)}]=0 \ ,
\eea
where we have taken into account that the product of $(N+1)$ 
translation ghosts automatically vanishes in a spacetime with 
$N$ dimensions, i.e. $s_{(1)}(\O^N_0)_{(N)}=0$.
The basis lemma~\cite{basis}
\bea
&&\sum_m~P_{(m)}\equiv 0~~~~~\NN_\x P_{(m)}=mP_{(m)}~~~~~
m=0,\ldots,N \non
&&~~~\Leftrightarrow~~~~~P_{(m)}\equiv 0~~~~~m=0,\ldots,N \ ,
\eea
states that a series in powers of $\x^\m$ is identically to zero 
if and only if each coefficient is equal to zero.
Therefore, the descent equations \equ{TOWER-2} devide into a set 
of descent equations according to the power of $\x^\m$.

\noindent
Since $s_{(0)}=s_g$, the descent equations in the $\x^\m$--sector 
$0$,
\bea
&&s_{(0)}(\O^0_N)_{(0)}+d(\O^1_{N-1})_{(0)}=0 \ , \non
&&s_{(0)}(\O^k_{N-k})_{(0)}+d(\O^{k+1}_{N-k-1})_{(0)}=0~~~,
~~~k=1,\ldots,N-1 \ , \non
&&s_{(0)}(\O^N_0)_{(0)}=0 \ ,
\label{x-0}
\eea
coincide with the descent equations of the gauge sector,
\bea
&&s_gQ^0_N+dQ^1_{N-1}=0 \ , \non
&&s_gQ^k_{N-k}+dQ^{k+1}_{N-k-1}=0~~~,~~~k=1,\ldots,N-1 \ , \non
&&s_gQ^N_0=0 \ ,
\label{TOWER_GAUGE}
\eea
and one has
\be
(\O^k_{N-k})_{(0)}=Q^k_{N-k}~~~,~~~k=0,\ldots,N \ .
\ee
A nontrivial solution for $Q^0_N$ represents an invariant
Lagrangian in a spacetime with $N$ dimensions, whereas a nontrivial
solution for $Q^1_{N-1}$ represents a candidate for an anomaly in a
spacetime with $N-1$ dimensions.
Generally, the descent equations in the $\x^\m$--sector $l$
can be collected as follows:
\bea
&&s_{(1)}(\O^{l-1}_{N-l+1})_{(l-1)}+d(\O^l_{N-l})_{(l)}=0 \ , \non
&&s_{(0)}(\O^k_{N-k})_{(l)}+s_{(1)}(\O^k_{N-k})_{(l-1)}
+d(\O^{k+1}_{N-k-1})_{(l)}=0~~~,~~~k=l,\ldots,N-1 \ , \non
&&s_{(0)}(\O^N_0)_{(l)}+s_{(1)}(\O^N_0)_{(l-1)}=0 \ .
\eea

\noindent
In order to solve the descent equations in the 
$\x^\m$--sectors $l\geq 1$ one defines the operator
\be
i_\x(dx^\m)=\x^\m \ ,
\ee
obeying the following algebraic relations:
\bea
[i_\x,d]=s_{(1)}~~~,~~~[i_\x,s_{(1)}]=0 \ .
\label{AREL}
\eea

\noindent
We continue with the discussion of the descent equations in the
$\x^\m$--sector $1$. 
Using the algebraic relation \equ{AREL}, the first descent 
equation in this sector,
\be 
s_{(1)}(\O^0_N)_{(0)}+d(\O^1_{N-1})_{(1)}=0,
\ee
becomes
\be
d\lt[(\O^1_{N-1})_{(1)}-i_\x(\O^0_N)_{(0)}\rt]=0 \ .
\ee
Since the $d$ cohomology is trivial~\cite{renorm,descent}, 
one obtains the result
\be
(\O^1_{N-1})_{(1)}=i_\x(\O^0_N)_{(0)}=i_\x Q^0_N \ .
\ee

\noindent
Analogously, the next $N-1$ descent equations
in the $\x^\m$--sector $1$ lead to
\be
(\O^k_{N-k})_{(1)}=i_\x(\O^{k-1}_{N-k+1})_{(0)}=
i_\x Q^{k-1}_{N-k+1}~~~,~~~k=2,\ldots,N \ .
\ee
One can show that the last descent equation in this
sector is then automatically fulfilled
\be
s_{(0)}(\O^N_0)_{(1)}+s_{(1)}(\O^N_0)_{(0)}=0 \ .
\ee
In the same manner one finds that the general nontrivial 
solution for the cocycles
of the descent equations \equ{TOWER-2}
can be expressed in terms of the general nontrivial solution 
for the cocycles of 
the descent equations in the gauge sector \equ{TOWER_GAUGE}
as follows
\bea
\O^k_{N-k}=\sum^k_{m=0}(\O^k_{N-k})_{(m)} 
=\sum^k_{m=0}\frac{1}{m!}\,(i_\x)^mQ^{k-m}_{N-k+m}~~~,~~~
k=0,\ldots,N \ .
\label{MAIN}
\eea
In particular the solution of the last descent equation in 
\equ{TOWER-2} becomes
\be
\O^N_0=\sum^N_{m=0}\frac{1}{m!}\,(i_\x)^mQ^{N-m}_m \ .
\ee
Since $\O^N_0$ collects all cocycles $Q^{N-m}_m$, $m=0,1,\ldots,N$,
of the descent equations in the gauge sector, one gets the 
following main result:
{\em The general nontrivial solution of the descent equations in 
the gauge sector is equivalent to the general nontrivial solution 
of the local equation}
\be
s\O^N_0=0 \ ,
\label{LOC_EQU}
\ee
{\em up to trivial terms.}

\noindent
Since $\O^N_0$ can be represented by an integrated parameter formula,
as it will be shown in the following, it is easier to solve the 
local equation \equ{LOC_EQU} instead of solving the tower of 
descent equations in the gauge sector.


\subsection{The decomposition}

In the preceding subsection it has been shown in \equ{MAIN}
that the general 
nontrivial solution for the cocycles $\O^k_{N-k}$
of the descent equations \equ{TOWER-2}
can be expressed in terms of the general nontrivial solution for 
the cocycles $Q^k_{N-k}$
of the descent equations in the gauge sector \equ{TOWER_GAUGE}.

\noindent
Defining the operator $\d$ as
\be
\d=dx^\m\frac{\6}{\6\x^\m} \ ,
\ee
the exterior derivative can be decomposed as a 
BRST commutator,
\be
[\d,s]=d \ .
\ee
Moreover, since the operator $\d$ and the exterior derivative
$d$ commute, $[\d,d]=0$,
an operator $\GG=\half\,[d,\d]$ is absent, in contrary 
to~\cite{decomp}.
Therefore, the algebra between the BRST operator, the exterior 
derivative and the $\d$--operator is closed,
\bea
[\d,s]=d~~~,~~~[\d,d]=0 \ .
\eea
Using the counting operators
\bea
\NN_{dx}=dx^\m\frac{\6}{\6(dx^\m)}~~~,~~~
\NN_\x=\x^\m\frac{\6}{\6\x^\m} \ ,
\eea
one obtains
\be
[\d,i_\x]=\NN_{dx}-\NN_\x \ ,
\ee
or generally, 
\bea
[\d,(i_\x)^m]=\sum^{m-1}_{l=0}(i_\x)^l\NN_{dx}(i_\x)^{m-l-1}
-\sum^{m-1}_{l=0}(i_\x)^l\NN_\x(i_\x)^{m-l-1} \ .
\eea

\noindent
The application of $\d$ to $\O^k_{N-k}$ gives
\bea
\d\O^k_{N-k}
\=\sum^k_{m=0}\frac{1}{m!}\,\d(i_\x)^m Q^{k-m}_{N-k+m} \non
\=(N-k+1)\sum^{k-1}_{p=0}
\frac{1}{p!}\,(i_\x)^pQ^{k-1-p}_{N-(k-1)+p} \non
\=(N-k+1)\O^{k-1}_{N-(k-1)} \ .
\eea
Therefore, one obtains the recursive relation
\be
\O^k_{N-k}=\frac{1}{N-k}\,\d\O^{k+1}_{N-k-1}~~~,~~~k=0,\ldots,N-1 \ .
\ee
The solution of the recursive relation is given by
\bea
\O^k_{N-k}=\frac{1}{(N-k)!}\,\d^{N-k}\O^N_0 \ ,
\eea
which is valid for $k=0,\ldots,N$.


\subsection{An integrated parameter formula}

In this subsection an integrated parameter formula for the general 
nontrivial solution of the local equation $s\O^N_0=0$
will be derived in a spacetime of dimension $N=2k-1$.
Expanding $\O^{2k-1}_0$ as a series in powers of $\x^\m$,
\bea
\O^{2k-1}_0\=Q^{2k-1}_0+i_\x Q^{2k-2}_1
+\ldots+\frac{1}{(2k-2)!}\,(i_\x)^{2k-2}Q^1_{2k-2} \non
\+\frac{1}{(2k-1)!}\,(i_\x)^{2k-1}Q^0_{2k-1} \ ,
\eea
and reducing the dimension of spacetime by one, the gauge anomaly in
$2k-2$ dimensions will be recovered as $Q^1_{2k-2}$.
Moreover, it will be shown that $Q^0_{2k-1}$ is the Chern--Simons 
term in $2k-1$ dimensions.

\noindent
In order to find the integrated parameter formula for $\O^{2k-1}_0$ 
we use the calculus of forms.
The $1$--form gauge field and the associated $2$--form field 
strength are given by
\be
A=A_\m dx^\m~~~,~~~F=\half\,F_{\m\n}\,dx^\m dx^\n=dA-iAA \ .
\ee
The Bianchi--identity reads
\bea
DF=dF-i[A,F]=0 \ ,
\eea
where $D=dx^\m D_\m$ denotes the covariant exterior derivative 
with respect to the $1$--form gauge field.
The BRST--transformation of the $1$--form gauge field becomes
\bea
sA\=-Dc+\x^\n\6_\n A = -dc+i\{A,c\}+\x^\n\6_\n A \non
\=-D(c+i_\x A)+i_\x F \ .
\eea
Introducing the shifted gauge ghost
\be
\ch=c+i_\x A \ ,
\ee
the BRST--transformations of the $1$--form gauge field
and the shifted gauge ghost are
\bea
sA\=-D\ch+i_\x F \ , \non
s\ch\=i\ch\ch+\Fh \ ,
\eea
with the ghost field strength
\be
\Fh=\half\,i_\x i_\x F \ ,
\ee
transforming according to 
\bea
s\Fh=i[\ch,\Fh] \ .
\eea
Defining a generalized covariant BRST--operator
\be
\Sh=s-i\ch \ ,
\ee
one gets
\be
\Sh\Fh=s\Fh-i[\ch,\Fh]=0 \ .
\ee
It follows a remarkable correspondence, which 
will be revealed by the following summary.
This correspondence is peculiar to the full BRST--operator $s$
including the translations.
\begin{itemize}
\item{The $1$--form gauge field corresponds to the shifted 
gauge ghost,
\be
A=A_\m dx^\m~~~,~~~\ch=c+i_\x A \ .
\label{A_HAT}
\ee
}
\item{The $2$--form field strength corresponds to the ghost 
field strength,
\bea
F\=\half\,F_{\m\n}\,dx^\m dx^\n=dA-iAA \ , \non
\Fh\=\half\,F_{\m\n}\,\x^\m\x^\n=s\ch-i\ch\ch \ .
\label{F_HAT}
\eea
}
\item{The covariant exterior derivative corresponds to the 
generalized covariant BRST--operator,
\be
D=d-iA~~~,~~~\Sh=s-i\ch \ .
\ee
}
\item{The Bianchi--identity
corresponds to the generalized covariant BRST--transforma-tion 
of the ghost field strength,
\be
DF=dF-i[A,F]=0~~~,~~~\Sh\Fh=s\Fh-i[\ch,\Fh]=0 \ .
\ee
}
\end{itemize}
The generalized covariant BRST--algebra is now given by
\bea
i\Sh^2=\Fh~~~,~~~iD^2=F~~~,~~~i\{\Sh,D\}=i_\x F \ .
\eea
Guided by the well--known derivation of the Chern--Simons 
formula~\cite{chern},
\be
(CS)^0_{2k-1}=k\,Tr\int^1_0dt\,AF^{k-1}(t) \ ,
\label{CS}
\ee
with
\bea
F(t)=dA(t)-iA(t)A(t)~~~,~~~A(t)=tA~~~,~~~0\leq t\leq 1 \ ,
\eea
and using the presented correspondence it will be easy to derive an 
integrated parameter formula for the general nontrivial solution of 
the local equation
\be
s\O^{2k-1}_0=0 \ .
\ee
We introduce the interpolating shifted gauge ghost
\be
\ch(t)=t\ch~~~,~~~0\leq t\leq 1 \ ,
\ee
with $\ch(0)=0$ and $\ch(1)=\ch$,
and the associated ghost field strength
\bea
\Fh(t)&=&s\ch(t)-i\ch(t)\ch(t) \ ,
\eea
with $\Fh(0)=0$ and $\Fh(1)=\Fh$.
Defining an interpolating generalized covariant BRST--operator
\be
\Sh_t=s-i\ch(t) \ ,
\ee
with $\Sh_0=s$ and $\Sh_1=\Sh$,
one gets the following identities
\bea
\frac{d\Fh(t)}{dt}=\Sh_t\ch~~~,~~~\Sh_t\Fh(t)=0 \ .
\eea
Therefore, in a spacetime with $2k$ dimensions one has
\bea
Tr(\Fh^k)\=Tr\lt(\Fh^k(1)-\Fh^k(0)\rt)
=Tr\int^1_0dt\,\frac{d}{dt}\,\Fh^k(t) \non
\=k\,Tr\int^1_0dt\,\frac{d\Fh(t)}{dt}\,\Fh^{k-1}(t)
=k\,Tr\int^1_0dt\,(S_t\ch)\Fh^{k-1}(t) \non
\=s\lt(k\,Tr\int^1_0dt\,\ch\,\Fh^{k-1}(t)\rt) \ .
\eea
Using the nilpotency of the BRST--operator and the 
fact that $Tr(\Fh^k)\neq 0$ in a spacetime with $2k$ dimensions, 
one concludes that $k\,Tr\int^1_0dt\,\ch\,\Fh^{k-1}(t)$ is 
nontrivial.
Since $Tr(\Fh^k)$ contains the product of $2k$ fermionic 
translation ghosts $\x^\m$, it follows:

\begin{quote}
{\em In a spacetime with $2k-1$ dimensions the general 
nontrivial solution of the local equation $s\O^{2k-1}_0=0$
can be represented by the integrated parameter formula}
\be
\O^{2k-1}_0=k\,Tr\int^1_0dt\,\ch\,\Fh^{k-1}(t) \ .
\label{IPF}
\ee
\end{quote}
Compare \equ{IPF} with \equ{CS} using the correspondence
\equ{A_HAT} and \equ{F_HAT}.

\noindent
Expanding $\O^{2k-1}_0$ as a series in powers of $\x^\m$,
\bea
\O^{2k-1}_0\=Q^{2k-1}_0+i_\x Q^{2k-2}_1
+\ldots+\frac{1}{(2k-1)!}\,(i_\x)^{2k-1}Q^0_{2k-1} \ ,
\eea
one gets the nontrivial solutions for the cocycles 
$Q^{2k-1-l}_l$ of the descent equations in the gauge sector.

\noindent
Finally, for the sake of clarity let us discuss in detail the 
special case $k=3$.
The tower of the descent equations in the gauge 
sector is then given by
\bea
&&s_gQ^0_5+dQ^1_4=0 \ , \non
&&s_gQ^1_4+dQ^2_3=0 \ , \non
&&s_gQ^2_3+dQ^3_2=0 \ , \non
&&s_gQ^3_2+dQ^4_1=0 \ , \non
&&s_gQ^4_1+dQ^5_0=0 \ , \non
&&s_gQ^5_0=0 \ .
\eea
In order to get the nontrivial solutions for the cocycles
$Q^k_{5-k}$,
one expands the integrated parameter formula \equ{IPF}
as a series in powers of $\x^\m$,
\bea
\O^5_0\=3\,Tr\int^1_0dt\,\ch\Fh(t)\Fh(t) \non
\=\four\,Tr\lt[(c+i_\x A)\,i_\x i_\x F\,i_\x i_\x F\rt] 
+\frac{i}{4}\,Tr\lt[(c+i_\x A)^3\,i_\x i_\x F\rt] \non
\-\frac{1}{10}\,Tr\lt[(c+i_\x A)^5\rt] \ .
\eea
After some calculations one obtains
\bea
\O^5_0\=-\frac{1}{10}\,Tr(ccccc) \non
\+i_\x\lt[Tr\lt(-\half\,ccccA\rt)\rt] \non
\+\half\,i_\x i_\x\lt[\half Tr(icccF-cccAA-ccAcA)\rt] \non
\+\frac{1}{6}\,i_\x i_\x i_\x\lt[Tr\lt(\frac{i}{2}\,(ccAF+cAcF+AccF)
-\half\,(ccAAA+cAcAA)\rt)\rt] \non
\+\frac{1}{24}\,i_\x i_\x i_\x i_\x\lt[Tr\lt(cFF
+\frac{i}{2}\,(cAAF+AcAF+AAcF)-\half\,cAAAA\rt)\rt] \non
\+\frac{1}{120}\,i_\x i_\x i_\x i_\x i_\x\lt[Tr\lt(AFF
+\frac{i}{2}\,AAAF-\frac{1}{10}\,AAAAA\rt)\rt] \ .
\eea
Since
\bea
\O^5_0\=Q^5_0+i_\x Q^4_1+\half\,i_\x i_\x Q^3_2
+\frac{1}{6}\,i_\x i_\x i_\x Q^2_3 \non
\+\frac{1}{24}\,i_\x i_\x i_\x i_\x Q^1_4
+\frac{1}{120}\,i_\x i_\x i_\x i_\x i_\x Q^0_5 \ ,
\eea
one concludes that the nontrivial solutions for the cocycles 
of the descent equations in the gauge sector are given by
\bea
Q^5_0\=-\frac{1}{10}\,Tr(ccccc) \ , \non
Q^4_1\=-\half\,Tr(ccccA) \ , \non
Q^3_2\=\half\,Tr(icccF-cccAA-ccAcA) \ , \non
Q^2_3\=\half\,Tr[i(ccAF+cAcF+AccF)-(ccAAA+cAcAA)] \ , \non
Q^1_4\=Tr\lt[cFF+\frac{i}{2}\,(cAAF+cAFA+cFAA)
-\half\,cAAAA\rt] \ , \non
Q^0_5\=Tr\lt(AFF+\frac{i}{2\,}AAAF-\frac{1}{10}\,AAAAA\rt) \ .
\eea
One sees that $Q^0_5$ is the Chern--Simons term in five dimensions
and $Q^1_4$ is the gauge anomaly in four dimensions~\cite{decomp}.


\section*{Appendix: The most general counterterm}

\setcounter{equation}{0}
\renewcommand{\theequation}{A.\arabic{equation}}

In this appendix the set of constraints 
\equ{INT_GHOST_EQU}--\equ{BRST_CC} 
will be solved leading to the result \equ{COUNTER_TERM}.
One starts with the most general perturbation
$\Sb_c=\Sb_c[A_\m,\,c,\,\rh^\m,\,\s]$
which has canonical dimension zero, ghost number zero and which is
invariant under the parity transformation,
\bea
\Sb_c\=Tr\ix\lt[k_1\6_\m A_\n(x)\6^\m A^\n(x)+
k_2\6_\m A_\n(x)\6^\n A^\m(x)\rt. \non
\+\lt.k_3\6_\m A_\n(x)A^\m(x)A^\n(x)+
k_4\6_\m A_\n(x)A^\n(x)A^\m(x)\rt. \non
\+\lt.k_5A_\m(x)A_\n(x)A^\m(x)A^\n(x)+
k_6A_\m(x)A_\n(x)A^\n(x)A^\m(x)\rt. \non
\+\lt.k_7\rh^\m(x)\6_\m c(x)+
k_8\rh^\m(x)A_\m(x)c(x)\rt. \non
\+\lt.k_9\rh^\m(x)c(x)A_\m(x)+
k_{10}\s(x)c(x)c(x)\rt] \ .
\eea
The functional derivative of $\Sb_c$ with respect to 
$c(x)$ is given by
\be
\frac{\d\Sb_c}{\d c(x)}=k_7\6_\m\rh^\m(x)-k_8\rh^\m(x)A_\m(x)-
k_9A_\m(x)\rh^\m(x)+k_{10}\lt[c(x),\s(x)\rt] \ .
\ee
From the integrated ghost equation \equ{INT_GHOST_EQU} 
follows that the functional derivative of $\Sb_c$ with 
respect to $c(x)$ has to be a total divergence.
Therefore, one has
\be
k_8=k_9=k_{10}=0 \ ,
\ee
and
\be
\frac{\d\Sb_c}{\d c(x)}=k_7\6_\m\rh^\m(x) \ .
\ee
The functional derivatives of $\Sb_c$ with respect to $\rh^\m(x)$ 
and $\s(x)$ are then
\bea
\frac{\d\Sb_c}{\d\rh^\m(x)}=k_7\6_\m c(x)~~~,~~~
\frac{\d\Sb_c}{\d\s(x)}=0 \ .  \label{SB_s}
\eea
Considering only the $\rh^\m$--dependence one gets
\be
\Sb_c=Tr\ix\lt[k_7\rh^\m(x)\6_\m c(x)+\ldots\rt] \ ,
\ee
where the low dots denote the terms which are independent 
of $\rh^\m$. Using
\bea
\SS_{\GO}\rh^\m(x)\=i\lt\{c(x),\rh^\m(x)\rt\}+\ldots \ , \\[3mm]
\SS_{\GO}A_\m(x)\=D_\m c(x) \ ,
\eea
the $\SS_{\GO}$--variation of
\be
\h{\S}=-k_7\,Tr\ix\,\rh^\m(x)A_\m(x)
\ee
becomes
\be
\SS_{\GO}\h{\S}=k_7\,Tr\ix\,\rh^\m(x)\6_\m c(x)+\ldots
\ee
and one gets
\be
\SS_{\GO}\h{\S}=\Sb_c+\ldots \ .
\ee
Hence the most general $\rh^\m$--dependence of the perturbation
$\Sb_c$ is trivial.
In order to find the most general nontrivial expression for
$\Sb_c$ one can thus assume
\be
\frac{\d\Sb_c}{\d\rh^\m(x)}=0 \ ,
\label{SB_r}
\ee
which fixes $k_7$ as
\be
k_7=0 \ ,
\ee
and therefore one also has
\be
\frac{\d\Sb_c}{\d c(x)}=0 \ .
\label{SB_c}
\ee
It remains
\bea
\Sb_c\=Tr\ix\lt[\6_\m A_\n(x)\lt(k_1\6^\m A^\n(x)
+k_2\6^\n A^\m(x)\rt.\rt. \non
\+\lt.\lt.k_3A^\m(x)A^\n(x)+k_4A^\n(x)A^\m(x)\rt)\rt. \non
\+\lt.A_\m(x)A_\n(x)\lt(k_5A^\m(x)A^\n(x)
+k_6A^\n(x)A^\m(x)\rt)\rt] \ .
\eea
The BRST--consistency condition becomes
\be
\SS_{\GO}(\Sb_c)=Tr\ix\,c(x)\DD(x)\Sb_c=0 \ ,
\ee
leading to the following set of constraints
\bea
k_1+k_2\=0~~~,~~~k_3+k_4=0~~~,~~~k_3-2ik_2=0 \ , \non
k_3+2ik_1\=0~~~,~~~k_4+2ik_2=0~~~,~~~k_4-2ik_1=0 \ , \non
ik_4-ik_3-4k_5\=0~~~,~~~2k_6-ik_3=0~~~,~~~2k_6+ik_4=0 \ ,
\eea
which has the solution
\bea
k_2\=-k_1~~~,~~~k_3=-2ik_1~~~,~~~k_4=2ik_1 \ , \non
k_5\=-k_1~~~,~~~k_6=k_1 \ .
\eea
Setting
\be
k_1=-\half\,k \ ,
\ee
one gets for the most general perturbation $\Sb_c$ of the action
in the tree approximation
\be
\Sb_c=-k\,\four\,Tr\ix\,F_{\m\n}(x)F^{\m\n}(x) \ .
\label{GENERAL_CT}
\ee
Using 
\be
\frac{\d\Sb_c}{\d A_\m(x)}=k\,D_\n F^{\n\m}(x) \ ,
\label{SB_A}
\ee
it easy to show that the perturbation \equ{GENERAL_CT} 
also obeys the Ward--identity describing the invariance 
under rigid gauge transformations,
\be
\h{\WW}_{rig}\Sb_c=
\ix\,i\lt[A_\m(x),\frac{\d\Sb_c}{\d A_\m(x)}\rt]=0 \ ,
\ee
and the Ward--identity describing the invariance under translations,
\be
\h{\PP}_\m\Sb_c=
Tr\ix\,\6_\m A_\n(x)\,\frac{\d\Sb_c}{\d A_\n(x)}=0 \ .
\ee
Thus $\Sb_c$ given by eq.\equ{GENERAL_CT} 
is the general solution of the set of constraints 
\equ{INT_GHOST_EQU}--\equ{BRST_CC}.




\begin{thebibliography}{99}


\bibitem{brst} C. Becchi, A. Rouet and R. Stora, 
              \annp{98}{76}{287} \\
              I.W. Tyutin, {\em Gauge Invariance in Field Theory 
              and Statistical Physics}, Lebedev Institute, 
              preprint FIAN no. {\bf 39} (1975);

\bibitem{qap} Y.--M. P. Lam,
              \pr{D6}{72}{2161} \\
              \pr{D7}{73}{2943}

\bibitem{renorm} O. Piguet and S.P. Sorella,\\
                 {\em Algebraic renormalization},\\
                 Lecture notes in physics, Springer (1995);

\bibitem{wz} J. Wess and B. Zumino, \pl{B37}{71}{95}

\bibitem{consis} M. Dubois-Violette, M. Talon and C.M. Viallet,
                 {\it Comm. Math. Phys.} 102 (1985) 105;\\ 
                 \pl{B158}{85}{231}\\
                 \aihp{44}{86}{103}\\
  M. Dubois-Violette, M. Henneaux, M. Talon and C.M. Viallet, 
                       {\it Phys. Lett.} B267 (1991) 81;\\
  M. Henneaux, \cmp{140}{91}{1}\\
  M. Dubois-Violette, M. Henneaux, M. Talon and C.M. Viallet, 
              {\it Phys. Lett.} B289 (1992) 361; 

\bibitem{cohom} G. Bandelloni,
                \jmp{27}{86}{2551} \\
                F. Brandt, {\it Structure of BRS-Invariant Local
                Functionals}, Ref. NIKHEF-H 93-21;

\bibitem{poinc} L. Bonora and P. Cotta-Ramusino, 
                \cmp{87}{83}{589} 

\bibitem{descent} F. Brandt, N. Dragon and M. Kreuzer, 
                  \np{B332}{90}{224} \\
                  \np{B340}{90}{187} 
                  

\bibitem{decomp} S. P. Sorella, \cmp{157}{93}{231}  \\
                 O. Moritsch, M. Schweda and S. P. Sorella,
                 \cqg{11}{94}{1225}

 
\bibitem{mo1} O. Moritsch, M. Schweda and T. Sommer, 
              \cqg{12}{95}{2059}

\bibitem{chern} F. Langouche, T. Sch\"ucker and R. Stora, 
                \pl{B145}{84}{342} \\
                R. Stora, 
                {\em Algebraic structure of chiral anomalies}; 
                lecture given at the GIFT seminar, 1-8 june 1985, 
                Jaca, Spain; 
                preprint LAPP-TH-143; \\
                J. Manes, R. Stora and B. Zumino, 
                \cmp{102}{85}{157} \\          
                T. Sch\"ucker, \cmp{109}{87}{167}

\bibitem{tree} C.N. Yang and R. Mills,
               \pr{96}{54}{191} \\
               O. Piguet and K. Sibold, \np{B253}{85}{517}

\bibitem{landau} A. Blasi, O. Piguet and S.P. Sorella,
                 \np{B356}{91}{154}

\bibitem{basis} F. Brandt, N. Dragon and M. Kreuzer, 
                \pl{B231}{89}{263} 
 




\end{thebibliography}
\end{document}